\documentclass[useAMS]{mn2e}
\usepackage{graphicx}
\usepackage{times}
\usepackage{amssymb}

\def\gsim { \lower .75ex \hbox{$\sim$} \llap{\raise .27ex \hbox{$>$}} }
\def\lsim { \lower .75ex \hbox{$\sim$} \llap{\raise .27ex \hbox{$<$}} }

\begin{document}

\title[Tidal debris in the solar neighbourhood]{Accretion relicts in
the solar neighbourhood: debris from $\omega$Cen's parent galaxy}

\author[Meza, Navarro, Abadi \& Steinmetz]{Andr\'es Meza,$^{1,4}$\thanks{Researcher of the
Academia Chilena de Ciencias 2004-2006.} Julio F. Navarro,$^{1}$\thanks{Fellow of CIAR and
of the J. S. Guggenheim Memorial Foundation.} Mario G. Abadi,$^{1}$\thanks{CITA National
Fellow, on leave from Observatorio Astron\'omico de C\'ordoba and CONICET, Argentina.} and
Matthias Steinmetz$^{2,3}$\thanks{David and Lucile Packard Fellow.}
\\
$^{1}$Department of Physics and Astronomy, University of Victoria, Victoria, BC V8P 5C2,
Canada\\
$^{2}$Astrophysikalisches Institut Potsdam, An der Sternwarte 16, Potsdam 14482, Germany\\
$^{3}$Steward Observatory, University of Arizona, Tucson, AZ 85721, USA\\
$^{4}$Departamento de F\'{\i}sica, Facultad de Ciencias F\'{\i}sicas y Matem\'aticas, 
Universidad de Chile, Casilla 487-3, Santiago, Chile}

\date{}
\pubyear{2004}
\maketitle
\label{firstpage}

\begin{abstract}
We use numerical simulations to investigate the orbital
characteristics of tidal debris from satellites whose orbits are
dragged into the plane of galactic disks by dynamical friction before
disruption. We find that these satellites may deposit a significant
fraction of their stars into the disk components of a galaxy, and use
our results to motivate the search for accretion relicts in samples of
metal-poor disk stars in the vicinity of the Sun. Satellites disrupted
on very eccentric orbits coplanar with the disk are expected to shed
stars in ``trails'' of distinct orbital energy and angular momentum
during each pericentric passage. To an observer located between the
pericenter and apocenter of such orbits, these trails would show as
distinct groupings of stars with low vertical velocity and a broad,
symmetric, often double-peaked distribution of Galactocentric radial
velocities. One group of stars with these characteristics stands out
in available compilations of nearby metal-poor stars. These stars have
specific angular momenta similar to that of the globular cluster
$\omega$Cen, long hypothesized to be the nucleus of a dwarf galaxy
disrupted by the Milky Way tidal field. In addition to their kindred
kinematics, stars in the $\omega$Cen group share distinct chemical
abundance characteristics, and trace a well-defined track in the
[$\alpha$/Fe] versus [Fe/H] plane, consistent with simple closed-box
enrichment models and a protracted star formation history. The
dynamical and chemical coherence of this group suggests that it
consists of stars that once belonged to the dwarf that brought
$\omega$Cen into the Galaxy. The presence of this and other ``tidal
relicts'' in the solar neighbourhood suggest an extra-Galactic origin
for the presence of nearby stars with odd kinematics and chemistry,
and implies that accounting for stars contributed by distinct
satellite galaxies may be crucial to the success of models of Galactic
chemical enrichment.
\end{abstract}


\section{Introduction}
\label{sec:intro}
Ever since Eggen, Lynden-Bell \& Sandage (1962, hereafter ELS) interpreted the
correlations between metallicity and kinematics of 221 stars in the solar
neighbourhood as evidence in favour of a monolithic collapse formation model of
the Milky Way, the mode of assembly of the Milky Way has been one of the main
goals of observational surveys of the chemistry, age, and kinematics of stars in
the Galaxy. Soon after the publication of ELS, it became clear that some data
provided challenges to the ELS interpretation, paving the way to more complex
formation scenarios for the Milky Way. For example, the properties of the
stellar halo, as traced by the globular cluster system, deviate from the trends
seen in the sample analyzed by ELS (Searle \& Zinn 1978, hereafter SZ) and have
long been viewed as consistent with a formation process that involved the merger
of a number of smaller subunits rather than the gradual collapse envisioned by
ELS.

The vigorous observational progress that followed the ELS and SZ papers
(recently reviewed by Freeman \& Bland-Hawthorn 2002) have led to the
development of a Milky Way formation scenario that borrows from the basic tenets
of both the ELS and SZ hypotheses. In this, the stellar spheroid is built
through a number of early mergers, whereas the stellar disk is regarded as the
outcome of the smooth, dissipative deposition (and transformation into stars) of
gas cooling more or less continuously off the intergalactic medium. Although the
disk is built smoothly in this scenario, it is not stationary, and may evolve as
a result of internal inhomogeneities such as spiral arms and molecular clouds,
as well as by the effect of minor mergers with external satellite
galaxies. Indeed, the presence of two distinct components in the Galactic disk
(the thin and thick disks, Gilmore \& Reid 1983) is widely ascribed to the
dynamical ``heating'' of an early thin disk by a satellite roughly $10$ Gyr ago
(Quinn \& Goodman 1986; Quinn, Hernquist \& Fullagar 1993; Robin et al. 1996;
Wyse 2004).

\begin{figure}
\begin{center}
\includegraphics[width=\linewidth,clip]{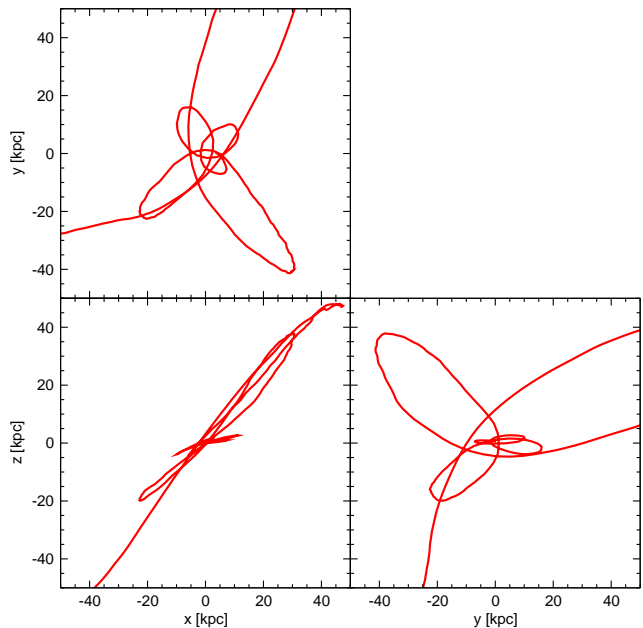}
\end{center}
\caption{Orbital trajectory of the satellite's self-bound core in a reference
frame where the main galaxy is at rest at the origin, and the $z$ axis is chosen
to coincide with the rotation axis of the disk component. The orbital plane is
initially seen approximately edge-on in the $x$-$z$ projection and is inclined
$\sim 40^\circ$ relative to the plane of the disk. As the orbit of the satellite
decays and approaches the disk, it is rapidly brought into the plane of the disk
before disruption. \label{figs:orbev}}
\end{figure}

Various versions of this scenario have long guided the interpretation
of newer datasets which, thanks to dedicated observing campaigns; to
the availability of accurate distance measurements from Hipparcos; as
well as to exquisite spectroscopy possible from 8-10m telescopes, have
expanded to include hundreds of stars with reliable three-dimensional
kinematics and detailed chemical abundance information. Most of these
stars, however, are in the immediate vicinity of the Sun, and
therefore the analysis suffers from the subtle and inevitable biases
that the Sun's location in the Galaxy brings about. The bias may be
modest if the stellar distribution function is a smooth function of
Galactic phase-space coordinates (May \& Binney 1986), but it may
imply severe limitations if a large number of coherent dynamical
groups---relicts from past accretion events---populate the Galaxy, as
these may well be under- or over-represented in the vicinity of the
Sun.

The relatively quiescent formation scenario of the Galaxy described
above---the Galactic disk is after all the dominant stellar component
of the Galaxy---has encouraged the interpretation of the abundance
patterns of nearby stars in terms of global collapse models where
enrichment proceeds according to the general scenario outlined by
Tinsley (1979). Their success in reproducing the correlations between
kinematics and chemistry of solar neighbourhood stars notwithstanding,
such models have traditionally been less successful at accounting for
the sizable scatter around the mean trends shown in the data, as well
as by the presence of abundance oddities in stars with otherwise
unremarkable kinematics (Carney et al.  1997; King 1997; Hanson et
al. 1998; Fulbright 2002).

\begin{figure}
\begin{center}
\includegraphics[width=\linewidth,clip]{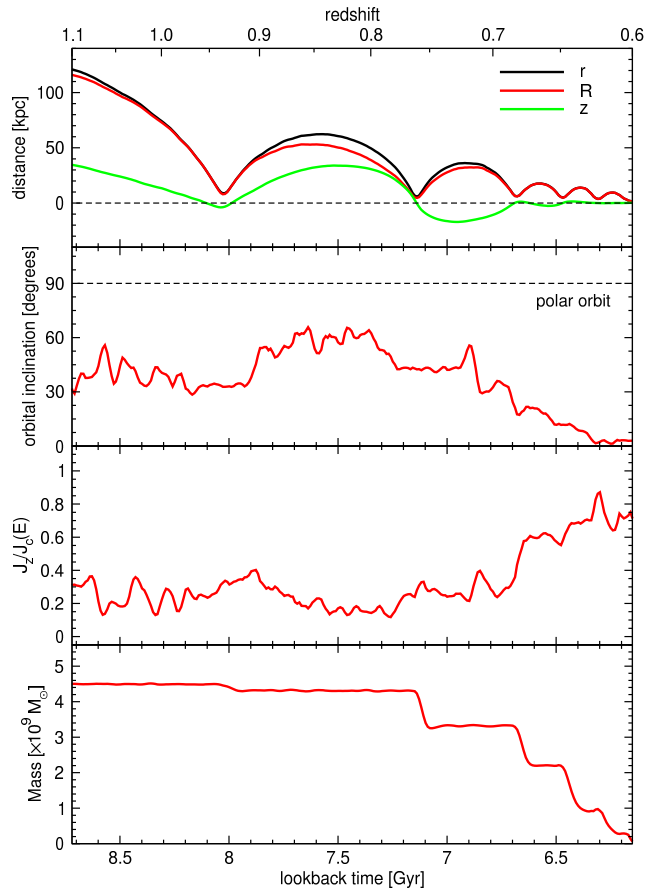}
\end{center}
\caption{
Evolution of the orbital parameters of the self-bound stellar core of the
satellite in the same reference frame chosen in Fig. \ref{figs:orbev}. Top panel
shows the evolution of the distance to the center of the main galaxy, $r$, the
distance to the rotation axis of the disk, $R$, and the vertical distance to the
plane of the disk, $z$, as a function of time. Second panel shows the evolution
of the instantaneous inclination of the orbital plane of the satellite relative
to the plane of the disk. Note that the satellite is brought onto an orbit
nearly coplanar with the disk before disruption. Third panel shows the evolution
of the circularity of the satellite's orbit, defined as the ratio between the
$z$-component of its angular momentum and that of a circular orbit of the same
binding energy. Note that the orbit becomes nearly circular before
disruption. Bottom panel shows the evolution of the self-bound stellar mass of
the satellite with time until disruption.\label{figs:orbpar}}
\end{figure}

From a cosmological perspective, a slow and quiescent buildup of the
Galaxy is at odds with the mode of assembly envisioned in currently
popular hierarchical models of structure formation such as the
$\Lambda$CDM scenario (Bahcall et al. 1999), where merging is expected
to have contributed significantly to the various populations of the
Galaxy (see, e.g., Steinmetz \& Navarro 2002). In particular, the
presence of an old, extended stellar disk component (such as that
hypothesized to be the progenitor of today's thick disk) is awkward to
accommodate in such models, since the dynamical fragility of stars in
disk-like orbits seems to preempt an active merging history.

\begin{figure*}
\begin{center}
\includegraphics[width=0.48\linewidth,clip]{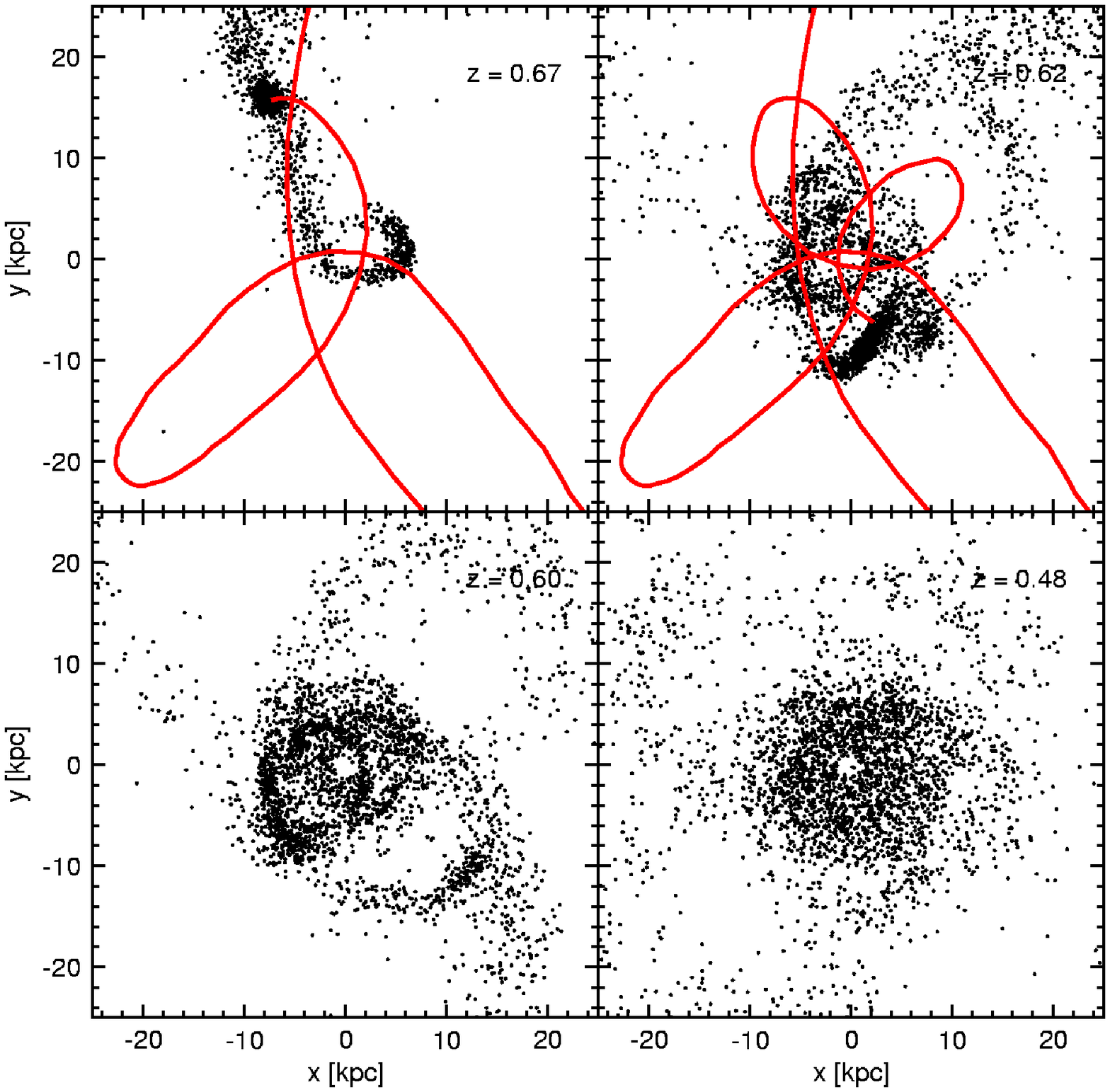}
\hspace{0.4cm}
\includegraphics[width=0.48\linewidth,clip]{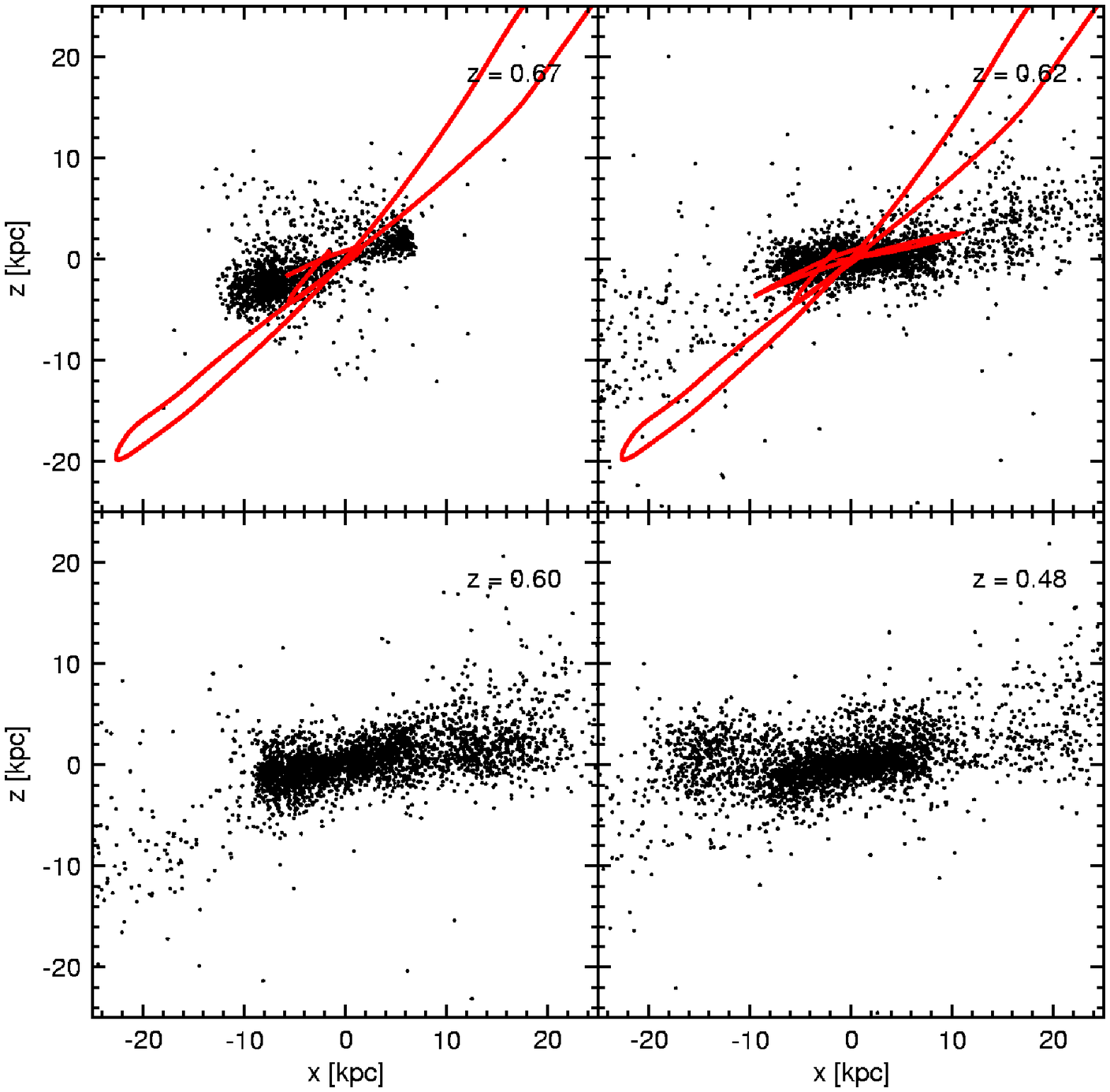}
\end{center}
\caption{
Evolution of the stellar component of the satellite seen ``face-on'' (left
panels) or ``edge-on'' (right panels) relative to the plane of the disk. The
curves in the top panels track the orbit of the self-bound core of the satellite
up until the time shown in each panel. The satellite starts being stripped of
stars after the pericentric passage that preceded the $z=0.67$ panel. Each
pericentric passage tears two tidal ``tails'' from the satellite, an inner one
more tightly bound and an outer one less tightly bound than the remaining
self-bound core. Although these tails maintain their identity in phase-space,
they quickly phase-mix, as the debris settles into a torus-shaped
configuration. Most stars are lost after the satellite is brought onto the plane
of the disk, and the symmetry axis of the torus is parallel to the rotation axis
of the disk. Thus, much of the debris ends up on eccentric, disk-confined orbits
(see text for further details).
\label{figs:posxyz}}
\end{figure*} 

Indeed, the presence of old stars in the disk components of the Galaxy
is often used to argue that the Galaxy has not been disturbed by
mergers throughout most of its life (see, e.g., Wyse 2004); clearly an
unusual assembly history in a universe where structure grows
hierarchically. The precise age of the Galactic disk components is
still somewhat controversial, given that ages of individual stars are
notoriously difficult to measure accurately, but the thin disk of the
Galaxy is known to contain a fair fraction of metal-poor (and
presumably quite old) stars. For example, roughly $16\%$ of stars in
the dynamically-unbiased catalog of metal-deficient ([Fe/H] $< -0.6$)
nearby stars compiled by Beers et al. (2000, hereafter B00) have thin
disk-like angular momenta.

Numerical simulations of hierarchical galaxy formation have suggested
a way to reconcile the presence of old, metal-poor disk stars with the
hectic merging activity envisioned in hierarchical formation
scenarios. Abadi et al. (2003a,b) report, for example, that $\sim15\%$
of dynamically-selected thin-disk stars in their cosmological
simulation of the formation of a disk galaxy are actually older than
the epoch of the galaxy's last major merger. Essentially all ($\gsim
90\%$) of these old stars are brought in at late times by satellite
galaxies whose orbits are circularized and dragged into the disk by
dynamical friction before disruption.

The core of satellites that disrupt after their orbits have been nearly
circularized may thus contribute stars to the thin disk component upon
disruption. In contrast, satellites that are disrupted before their orbits are
substantially eroded by dynamical friction (or whose orbits are roughly polar to
the disk) will contribute most of their stars to the spheroidal component. More
massive satellites are likely to see their orbits more severely affected by
dynamical friction before disruption (they are typically denser and thus more
resilient to disruption), and are therefore likely to shed their stars onto more
rotationally supported orbits than those of less massive companions more prone
to disruption.

This suggests that broad correlations between metallicity and rotation
may arise without the need for progressive in-situ enrichment of gas
cooling and settling into the disk. Indeed, most metal poor stars in
the solar neighbourhood may actually be the overlapping debris of
numerous satellites disrupted in the past. In such scenario, the trend
for rotational support to increase with metallicity would just reflect
the mass-metallicity relation of their progenitor galaxies, coupled
with the more efficient orbital decay of massive satellites in the
potential of the Galaxy.

If this scenario is correct, it should be possible to identify at
least some of the relicts of such accretion events as coherent groups
in phase space. A number of such kinematically-distinct associations
have been recognized in the stellar halo; for example, the Sagittarius
dwarf (Ibata, Gilmore \& Irwin 1994); the retrograde rotation of halo
stars in the direction of the North Galactic Pole (Majewski 1992); and
the identification of substructure in the outer halo by Helmi et
al. (1999).

\begin{figure*}
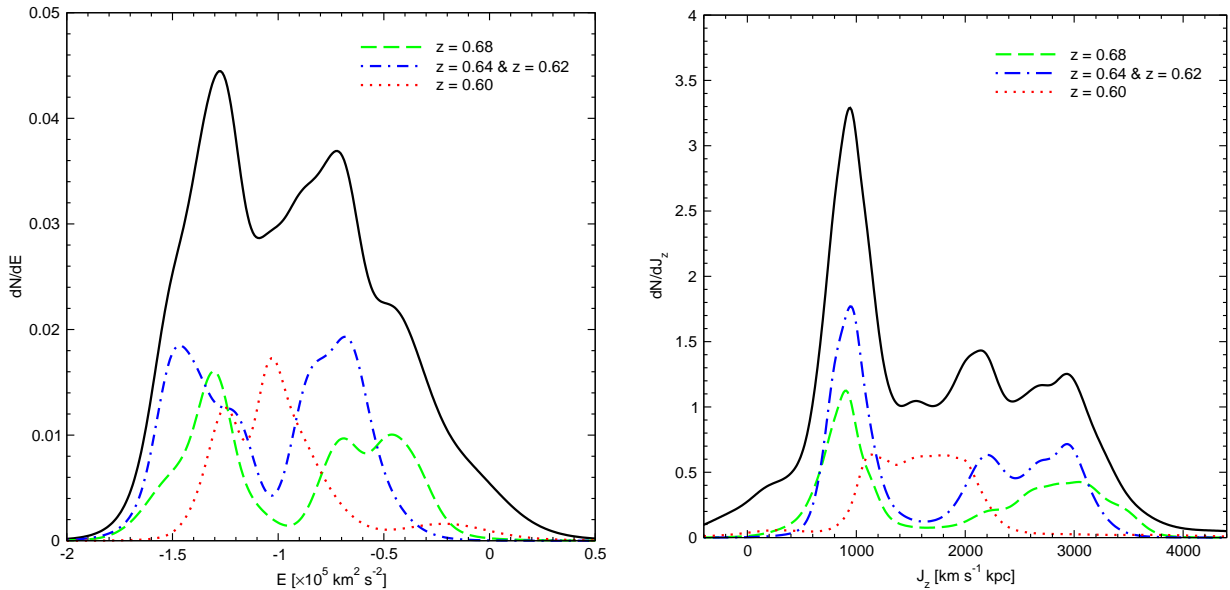

\begin{center}
\includegraphics[width=0.45\linewidth,clip]{f4a.eps}
\hspace{0.4cm}
\includegraphics[width=0.437\linewidth,clip]{f4b.eps}
\end{center}
\caption{
Specific energy ($E$) and $z$-angular momentum ($J_z$) distribution of satellite
stellar debris soon after full disruption ($z\sim 0.60$). The thick upper curve
corresponds to all satellite stars; the bottom curves show the $E$ and $J_z$
distribution of particles stripped from the galaxy during the last three
pericentric passages: dashed curves correspond to stars stripped at $z=0.68$,
dot-dashed curves at $z=0.64$ and $z=0.62$, and dotted curves at $z=0.60$. Stars
stripped at each pericentric passage form long-lived, dynamically distinct
groups in $E$-$J_z$ space. \label{figs:histos}}
\end{figure*} 

Accretion onto the disk has also been recently recognized.  One example is
provided by the ``Monoceros ring'' of stars in the outer disk of the Galaxy,
which has been interpreted as material recently stripped from a dwarf galaxy
that may still remain hidden in the disk (Yanny et al. 2003; Newberg et
al. 2002; Helmi et al. 2003; Ibata et al. 2003; Rocha-Pinto et al.  2003; Crane
et al. 2003; Martin et al. 2004). Further examples are provided by the
``Arcturus stream''; a group of stars at the apocenter of their eccentric, disk
confined orbits that has been interpreted as debris from the disruption of a
satellite in the plane of the Galactic disk (Eggen 1971; Navarro, Helmi \&
Freeman 2004); as well as by the discovery of an unexpected, slowly-rotating
component above and below the plane of the disk (Gilmore, Wyse
\& Norris 2002).

We examine here the possibility that other solar-neighbourhood stars
may have originated from material stripped from disrupted satellites.
As with the analysis of the ``Arcturus group'', we shall focus on
stars with orbits confined to relatively small excursions from the
plane of the disk (but not necessarily on circular orbits) and use
numerical simulations to motivate and guide the search for common
orbital characteristics in catalogs of nearby metal-deficient
stars. 

We describe briefly the numerical simulations and present our
main results in Section \ref{sec:numexp}. We use these to motivate the
analysis of samples of nearby stars with accurate kinematics and
chemical abundance measurements in Section \ref{sec:omcen}. We
conclude with a brief summary and discussion of our main conclusions
in Section \ref{sec:conc}.

\section{Tidal Debris in Cosmological Simulations}
\label{sec:numexp}

\subsection{Numerical experiments}

We analyze the accretion and disruption of a satellite galaxy in the
simulation of the formation of a disk galaxy presented by Abadi et
al. (2003a,b).  The simulation follows self-consistently the evolution
in a $\Lambda$CDM universe of a small region surrounding a target
galaxy, excised from a large periodic box and resimulated at higher
resolution preserving the tidal fields from the whole box. The
simulation includes the gravitational effects of dark matter, gas and
stars, and follows the hydrodynamical evolution of the gaseous
component using the Smooth Particle Hydrodynamics (SPH)
technique. Dense, cold gas is allowed to turn into stars at rates
consistent with the empirical Schmidt-like law of Kennicutt
(1998). The energetic feedback of evolving stars is included as a
heating term on the surrounding gas, but its effectiveness in
curtailing star formation is low. The transformation of gas into stars
thus tracks closely the rate at which gas cools and condenses at the
center of dark matter halos. Details of the simulation are given in
Abadi et al. (2003a,b).

As discussed by these authors, at $z=0$ the main galaxy resembles more an
early-type spiral than the Milky Way. Although the luminosity of the disk
component ($L_{\rm disk} \sim 2 \times 10^{10} \, L_{\odot}$) and the circular
speed at $R=10$ kpc ($\sim 250$ km s$^{-1}$) are comparable to those of our
Galaxy we emphasize that this is {\it not} a model of the Milky Way, and that
our main goal is to understand {\it qualitatively} rather than quantitatively
the nature of debris from tidally disrupted satellites.

\subsection{Satellite orbital evolution}

We choose to illustrate the analysis with the same satellite used by Helmi et
al. (2003) in their discussion of the various formation scenarios of the
``Monoceros ring''. This is a $4.6 \times 10^9\, M_{\odot}$ satellite which is
accreted and disrupted at $z\sim 0.6$ by the main galaxy. The satellite turns
around at $z\sim 1.2$ from a radius of $140$ (physical) kpc, and first
approaches the galaxy at $z=0.94$, when it reaches a pericentric radius of $7.6$
(physical) kpc. Its highly eccentric orbit is inclined by $\sim 40^\circ$
relative to that of the disk of the galaxy. These orbital parameters evolve
quickly, as dynamical friction breaks the satellite's orbit and steadily reduces
its apocentric radius.

Fig. \ref{figs:orbev} shows the trajectory of the satellite in three orthogonal
projections, chosen so that the $z$ axis coincides with the rotation axis of the
disk of the main galaxy, which is located at the origin of the coordinate
system. This trajectory follows the center of mass of the self-bound stellar
core of the satellite, computed by iteratively removing escapers using the
procedure outlined in detail by Hayashi et al. (2003). For short, we shall use
``satellite'' to denote this self-bound group of stars that track the surviving
core of the satellite.

\begin{figure*}
\begin{center}
\includegraphics[width=0.85\linewidth,clip]{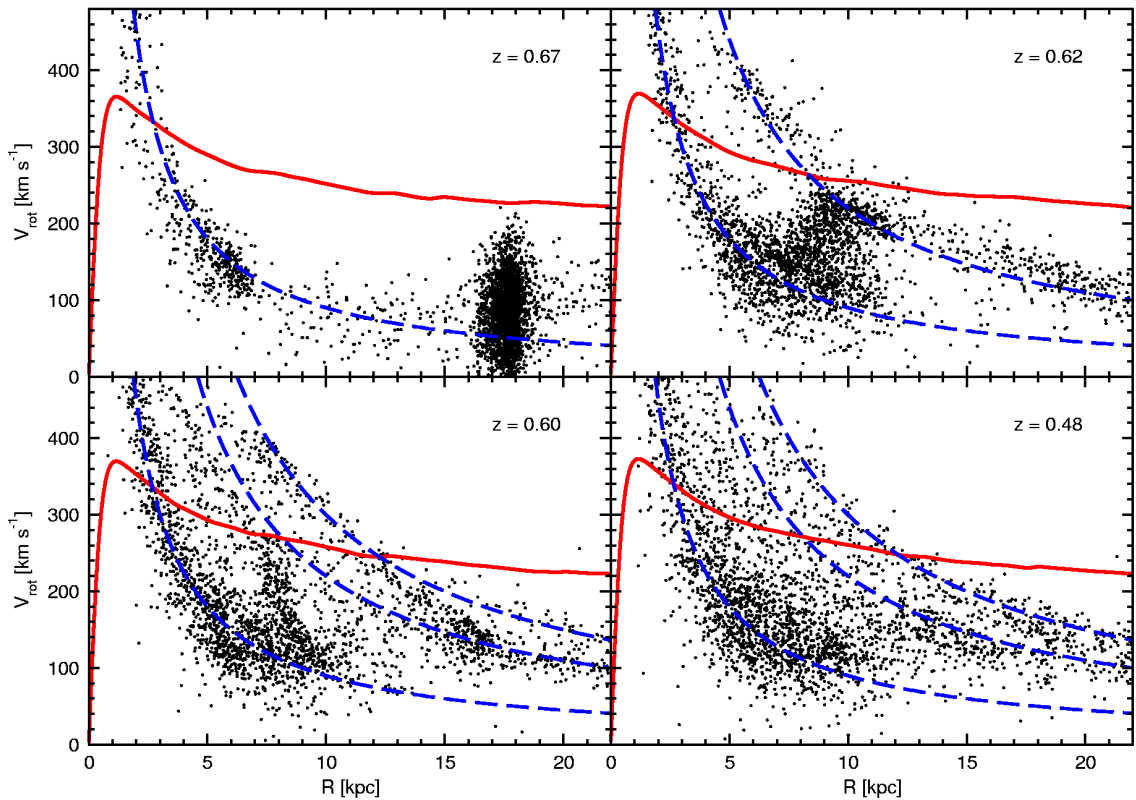}
\end{center}
\caption{
Evolution of satellite stars in the $R$-$V_{\rm rot}$ plane. $R$ is the distance to
the rotation axis of the disk, and $V_{\rm rot}=J_z/R$ is the rotation speed about
that axis. Dashed curves are lines of constant (specific) angular momentum that
mark the three most prominent ``peaks'' in the $J_z$ distribution shown in
Fig. \ref{figs:histos}. From top to bottom, these correspond to $J_z=900$,
$2200$, and $3000$ km s$^{-1}$ kpc. Solid curve is the circular velocity of the
main galaxy, $V_c(R)$, measured on the plane of the disk.
\label{figs:vphi}}
\end{figure*} 

The orbit of the satellite decays steadily, gradually becoming more circular and more
closely aligned with the plane of the disk. This is shown in Fig. \ref{figs:orbpar}, where
we show the evolution of the satellite's distance to the galaxy (top panel); the inclination
of the orbit relative to the disk (second panel); the orbital circularity\footnote{The
circularity $J_z/J_c(E)$ is defined as the ratio between the orbital angular momentum and
that of a circular orbit with the same binding energy, $E$.} (third panel); and the
self-bound stellar mass of the satellite (bottom panel).

In the last few orbits the satellite's core decays to an almost circular orbit
coplanar with the disk. Although the orbital energy (as measured by the
satellite's apocenter) declines gradually throughout the evolution, the
inclination and circularity of the satellite change noticeably as the satellite
approaches the disk of the main galaxy, which extends out to $\sim 10$-$15$ kpc
at this redshift. The fast decay into the plane seen after $z\sim 0.75$ is
accompanied by rapid circularization of the orbit of the remaining self-bound
core (Fig. \ref{figs:orbpar}), and is consistent with the results of previous
work on satellite decay in disk-dominated or flattened systems (see, e.g., Quinn
\& Goodman 1986; Pe\~narrubia, Kroupa \& Boily 2002, and references therein).

\subsection{Satellite debris}

The stellar component of the satellite is disrupted mainly during the three last
pericentric passages, between $z=0.7$ and $0.6$. The self-bound mass of the
satellite core drops to zero after $z\sim 0.6$, and the satellite debris quickly
phase-mixes into a torus-shaped configuration, as shown in
Fig. \ref{figs:posxyz}. The well-mixed, relaxed configuration of the debris at
late times hides the presence of several well-defined, dynamically coherent
groups of well-defined (specific) binding energy ($E$) and angular momentum
($J_z$). This is shown in Fig. \ref{figs:histos}, where several distinct groups
of particles are clearly apparent in the $E$ and $J_z$ histograms after
disruption.

\begin{figure*}
\begin{center}
\includegraphics[width=0.85\linewidth,clip]{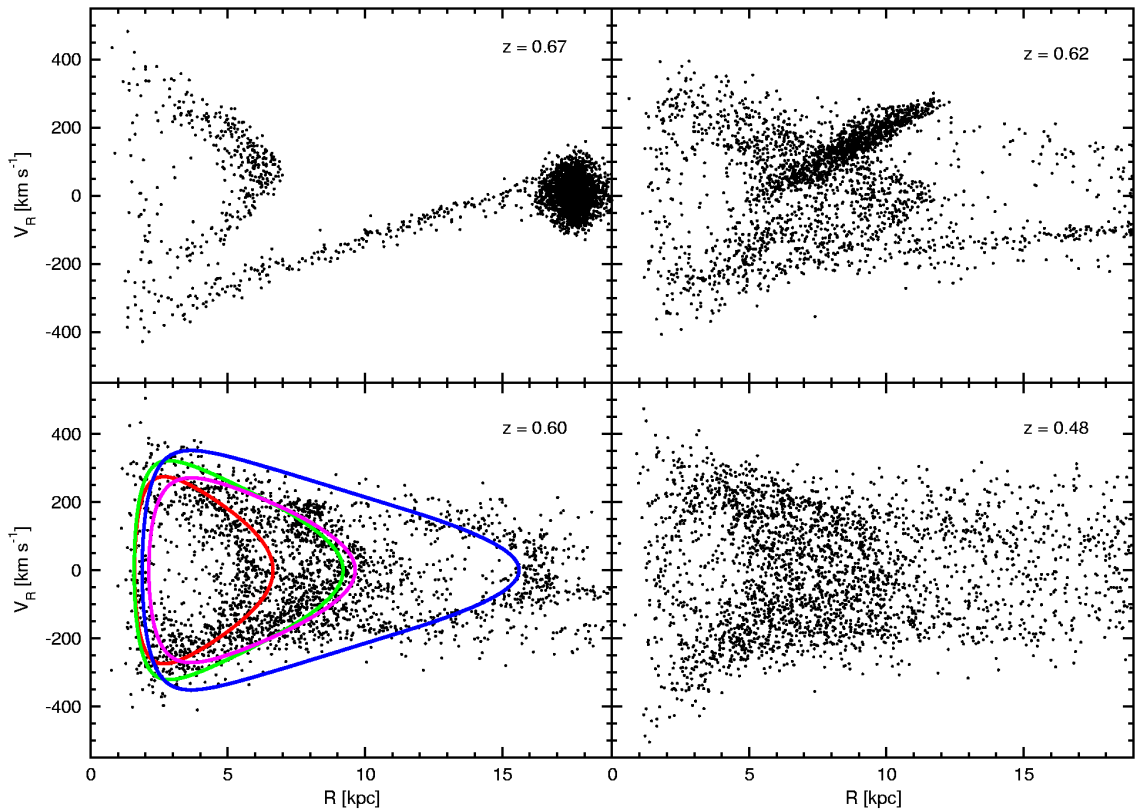}
\end{center}
\caption{
Evolution of satellite stars in the $R$-$V_{R}$ plane. $R$ is the distance to
the rotation axis of the disk, and $V_{R}$ is the galactocentric radial
velocity. The solid curves are lines of constant (specific) energy and angular
momentum that mark some of the characterisitic ``peaks'' in the $J_z$ and $E$
distributions shown in Fig. \ref{figs:histos}. From the inside out, ($E$, $J_z$)
for each curve is ($-1.5\times 10^5$, $900$); ($-1.3\times 10^5$, $950$); ($-1.25\times 10^5$, $1150$); and
($-1.0\times 10^5$, $1150$), respectively. $E$ is in units of km$^2$ s$^{-2}$ and $J_z$ is in
km s$^{-1}$ kpc.
\label{figs:vr}}
\end{figure*} 

The ``peaks'' in these histograms may be traced to particles lost during each of
the last few three pericentric passages prior to full disruption. This is
illustrated by the dashed curve in Fig. \ref{figs:histos}, which correspond to
stars lost at $z\approx 0.68$. The dot-dashed curve corresponds to stars lost
during the last two pericentric passages, at $z\approx 0.64$ and $0.62$,
respectively. In general, the satellite sheds two groups of particles at each
pericentric passage, corresponding roughly to the ``inner'' and ``outer'' tidal
arms seen in the $z=0.67$ panel of Fig. \ref{figs:posxyz}, so we expect several
distinct groups of particles in Fig. \ref{figs:histos}. Because of overlaps
between groups, as well as the intrinsic spread in the $E$ and $J_z$
distributions due to the finite velocity dispersion of the satellite, roughly
three prominent groups are clearly identifiable by their distinct energy and
angular momentum after disruption.

The evolution of these groups in position-velocity space is shown in
Figs.~\ref{figs:vphi} and ~\ref{figs:vr}, which show the rotation ($V_{\rm
rot}=J_z/R$) and radial ($V_R$) velocity as a function of their distance ($R$)
to the rotation axis of the disk. The solid curve in Fig. \ref{figs:vphi} shows
the circular velocity on the plane of the disk of the main galaxy. Dashed
hyperbolae in this figure are lines of constant specific angular momentum which
indicate the location of three prominent ``peaks'' in the $J_z$ distribution
shown in Fig. \ref{figs:histos}.

These groupings are also seen in the $V_R$-$R$ plane, as shown in
Fig. \ref{figs:vr}. Solid line curves in the $z=0.60$ panel of this
figure indicate the average trajectory of particles with various
energies and angular momenta. The innermost and outermost loops trace
orbits with ($E$/km s$^{-2}$, $J_z$/km s$^{-1}$ kpc) of order
($-1.0\times 10^5$, $1150$) and ($-1.5 \times 10^5$, $1150$),
corresponding roughly to two of the peaks in the distribution of
stripped stars seen in Figure~\ref{figs:histos}. These groupings in
$E$-$J_z$ space are expected to be long lived, provided that the
galactic potential remains relatively quiescent afterwards.

\subsection{Debris identification strategy}

As discussed by Abadi et al. (2003a,b), the disruption of satellites on coplanar
orbits provides a mechanism for adding stars of various ages and metallicities
to the disk components of a galaxy. These authors concentrate their analysis on
the debris that contributes to the thin disk, and argue that accretion events
are possibly responsible for the majority of old thin-disk stars in the solar
neighbourhood. Satellites like the one we analyze here do not contribute
substantially to the thin disk, as few particles in the debris of this satellite
end up on nearly circular orbits. However, most of the debris is confined to a
plane roughly coincident with that of the disk of the galaxy, as shown in
Fig. \ref{figs:posxyz}, and might therefore be considered part of a thick disk
component.

\begin{figure}
\begin{center}
\includegraphics[width=\linewidth,clip]{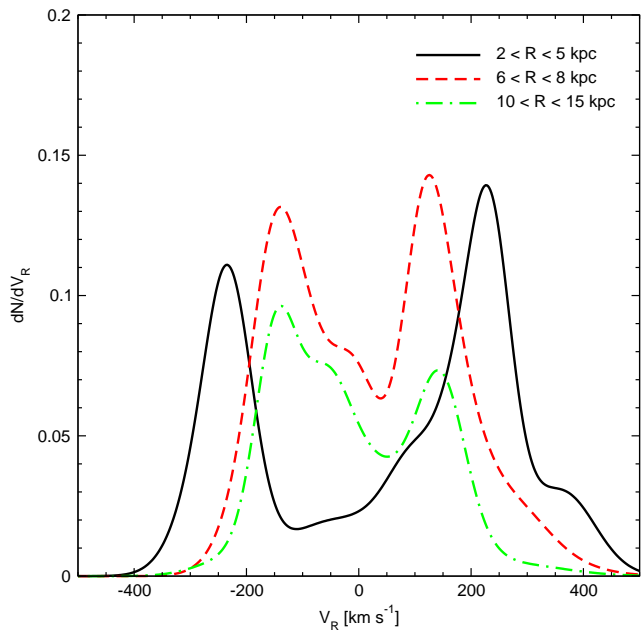}
\end{center}
\caption{Distribution of galactocentric radial velocities of the satellite
debris measured after full disruption ($z=0.6$) at various radii. The solid
curve is the histogram measured in the range ($2$, $5$) kpc, the dashed curve
corresponds to the range ($6$, $8$) kpc, and the dashed curve to ($10$, $15$)
kpc. These locations are chosen to bracket the apocentric radii corresponding to
the three most prominent ``peaks'' in the distribution of specific energies
shown in Fig. \ref{figs:histos}, and are intended to illustrate the broad,
symmetric, ``double-humped'' velocity distribution measured by observers in the
disk at radii intermediate between the characteristic apocentric radii of the
debris.
\label{figs:histvr}}
\end{figure} 

To an observer in the disk, the local properties of the satellite debris will
depend significantly on the position of the observer relative to the pericenter
and apocenter of each of the groups mentioned above. Local detection of these
groupings as recognizable features against the background of Galactic stars
would be easier near apocenter, since stars tend to crowd there during their
orbits.  A possible debris detection strategy would thus be to search samples of
nearby stars for groups with: (i) modest vertical motions, (ii) negligible net
Galactocentric radial velocity, and (iii) common angular momentum, and to probe
them for corroborating traits suggestive of a common origin.

One example of such approach is provided by the ``Arcturus group'' reported by
Navarro et al. (2004). These authors argue that the significant excess of disk
stars with angular momentum similar to Arcturus' seen in the B00 catalog,
together with other, more indirect evidence (such as the tight correlation in
the [$\alpha$/Fe] versus [Fe/H] plane, and the discovery of an unexpected
population with similar rotation speed above and below the Galactic plane by
Gilmore et al. (2002)), constitute persuasive evidence for the extra-Galactic
origin of the bright star Arcturus and of other members of its namesake group.

On the other hand, for observers located {\it between} the apocenter and
pericenter of a well-mixed group the debris would show locally as groups of
stars with a rather broad, symmetric (and at times ``double-humped'')
distribution of Galactocentric radial velocities. This is illustrated in
Fig. \ref{figs:histvr}, where we show the $V_R$ histogram of stars chosen at
various radii across the disk after disruption ($z=0.6$). These radii are chosen
to lie {\it between} the apocenters of the most prominent groups identified in
Fig. \ref{figs:histos}.  The relatively symmetric, double-peaked distribution of
Galactocentric radial velocities is a telltale sign of a past accretion event
that left a substantial fraction of stars on orbits more energetic than the
observer's. We shall use this result below to motivate the search for coherent
dynamical structures in samples of nearby stars that may have originated in the
disruption of former satellites of the Milky Way.

\section{Tidal Debris in the solar neighbourhood}
\label{sec:omcen}

Given the clear trend for the metallicity of galaxies in the Local
Group to increase with luminosity (van den Bergh 1999), signatures of
the accretion and disruption of dwarf galaxies are expected to be more
prevalent in samples of metal-poor stars. B00 provide a catalog of
metal-deficient stars in the vicinity of the Sun, compiled with the
specific aim of minimizing kinematic selection biases. This is the
largest sample of metal-poor ([Fe/H] $\lsim -0.6$) stars with
available Fe abundances, distances, and radial velocities. Proper
motions are also available for many of these stars, making it a good
sample to search for substructure (see, e.g., Chiba \& Beers 2000;
Brook et al. 2003).

The top dot-dashed histogram in Fig.~\ref{figs:vphihist} shows the distribution
of Galactocentric rotation speeds for all stars in the B00 sample. The full
sample is dominated by the halo and the canonical ``thick disk'' component.  A
decomposition of the $V_{\rm rot}$ distribution into three Gaussians
representing the ``canonical'' components of the Galaxy is shown in
Fig. \ref{figs:vphihist}. From left to right, the halo, thick disk, and thin
disk components are assumed to have ($\langle V_{\rm rot} \rangle$, $\sigma$)
equal to ($0$,$110$), ($160$,$50$) and ($220$,$25$) km s$^{-1}$, and are found
to contribute $56\%$, $28\%$, and $16\%$ of the sample, respectively.

The halo distribution is normalized to match the number of counter-rotating
stars, whereas the relative contributions of the thick and thin disks are chosen
to match the $V_{\rm rot}$ distribution of the remaining co-rotating stars. The
fit obtained is shown by the thick solid curve in Fig.~\ref{figs:vphihist}.
Most random realizations are unable to account for the ``excess'' of stars
labelled as the ``Arcturus group'' at $V_{\rm rot}\sim 90$ km s$^{-1}$; only
$16$ out of $1000$ realizations give a number of stars comparable to what is
observed (Navarro et al. 2004).

\subsection{The $\omega$Cen group}

\begin{figure}
\begin{center}
\includegraphics[width=\linewidth,clip]{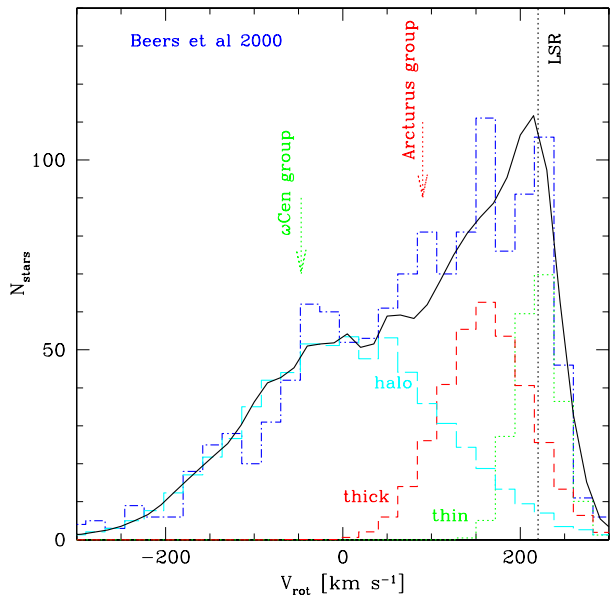}
\end{center}
\caption{
Rotational velocity distribution of all stars in the dynamically unbiased
catalog of metal-poor ([Fe/H] $\lsim -0.6$) stars of Beers et al. (2000) (dashed
histogram), decomposed into three canonical Galactic (Gaussian) components:
halo, thick disk, and thin disk, as labelled. The solid curve denotes the sum of
all three components. Note the excess of stars lagging the local standard of
rest (LSR) rotation by $\sim 120$ km s$^{-1}$, coincident with the Arcturus group, as
well as that for mildly retrograde orbits labelled as the $\omega$Cen group. See
text for full details.
\label{figs:vphihist}}
\end{figure}

Of similar significance is the excess of stars on slightly retrograde orbits
($-50$ km s$^{-1} < V_{\rm rot} < 0$ km s$^{-1}$) labelled as the ``$\omega$Cen
group'' in Fig.~\ref{figs:vphihist}. Stars in this group have specific angular
momenta similar to that of the globular cluster $\omega$Cen, the most luminous
and unusual of the Milky Way globular clusters.  Although the typical
metallicity of stars in $\omega$Cen is low (its distribution peaks at [Fe/H]
$\sim -1.6$; see Smith 2004 for a review), there is convincing evidence for a
long and complex star formation and enrichment history in the structure of the
giant branch of cluster stars (Norris, Freeman \& Mighell 1996; Lee et
al. 1999), and in the unusual s-process element abundances, which point to the
retention of elements fabricated in AGB stars (Norris \& Da Costa 1995).

These properties strongly suggest that the cluster did not form on its present
orbit, which is rather bound (apocenter $\sim 6.2$ kpc), confined to the disk
($z_{\rm max} \sim 1$ kpc), and mildly retrograde (Dinescu, Girard \& van Altena
1999). This orbit implies frequent passages through the disk, and may only be
reconciled with significant gas retention and repeated episodes of star
formation if $\omega$Cen formed elsewhere and its orbit has decayed
substantially over time. Efficient decay requires more mass than presently
attached to the cluster, implying that $\omega$Cen was likely much more massive
in the past. Indeed, it has long been hypothesized that it was the nucleus of a
rather massive and dense dwarf accreted and disrupted in the tidal field of the
Milky Way (Freeman 1993).

Although the apocenter of $\omega$Cen's orbit lies at present within the solar circle, a
substantial fraction of the stars of its parent galaxy may have been shed into orbits of
higher energy that now intersect the solar neighbourhood. This has motivated several authors
to search for debris from $\omega$Cen's parent galaxy in catalogs of nearby metal-poor
stars. Dinescu (2002), for example, argues that the retrograde signature noted in
Fig. \ref{figs:vphihist} for stars with $V_{\rm rot} \sim -50$ km s$^{-1}$ is most
prevalent in samples with metallicities restricted to $-2.0<$ [Fe/H] $<-1.5$; a range that
overlaps the metallicity distribution of $\omega$Cen stars.

Tidal debris tends to settle into a structure more vertically extended than the
thin disk of the Galaxy, and therefore it may be detectable {\it in situ} above
and below the plane of the Galaxy.  Mizutani, Chiba \& Sakamoto (2003; see also
Chiba \& Mizutani 2004) have recently argued that this may explain the presence
of a high-velocity ($V_{\rm los} \sim 300$ km s$^{-1}$) peak in the
line-of-sight velocities of stars in the direction against Galactic rotation
recently presented by Gilmore et al. (2002).

\begin{figure}
\begin{center}
\includegraphics[width=\linewidth,clip]{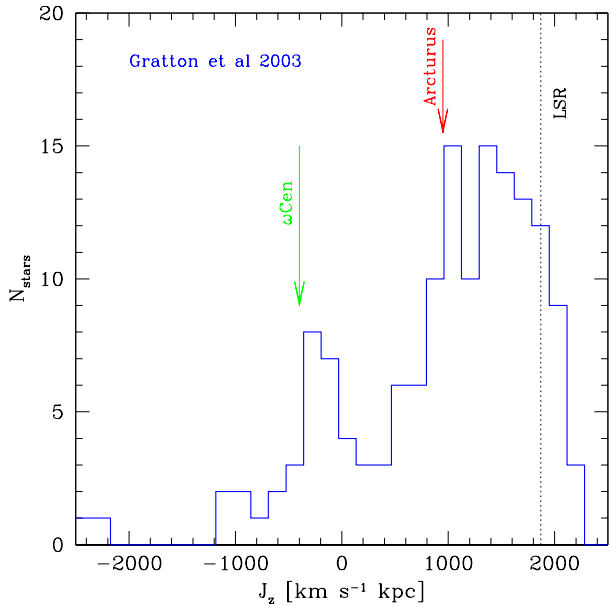}
\end{center}
\caption{
Distribution of specific angular momenta of all stars in the metal-poor
compilation of Gratton et al. (2003). These $\sim 150$ stars have accurate
kinematics and distances, as well as reliable element-to-element abundance
ratios. The excess of stars corresponding to the Arcturus and $\omega$Cen groups
noted in the Beers et al. (2000) catalog (Fig. \ref{figs:vphihist}) is also
clearly seen here in this independent compilation.
\label{figs:jzgrat}}
\end{figure} 

Our analysis is complementary to that of these authors, but focuses instead on the
Galactocentric radial velocity distribution of disk-confined stars with angular momenta
similar to $\omega$Cen's, as well as on the detailed abundance patterns of such stars.

\begin{figure}
\begin{center}
\includegraphics[width=\linewidth,clip]{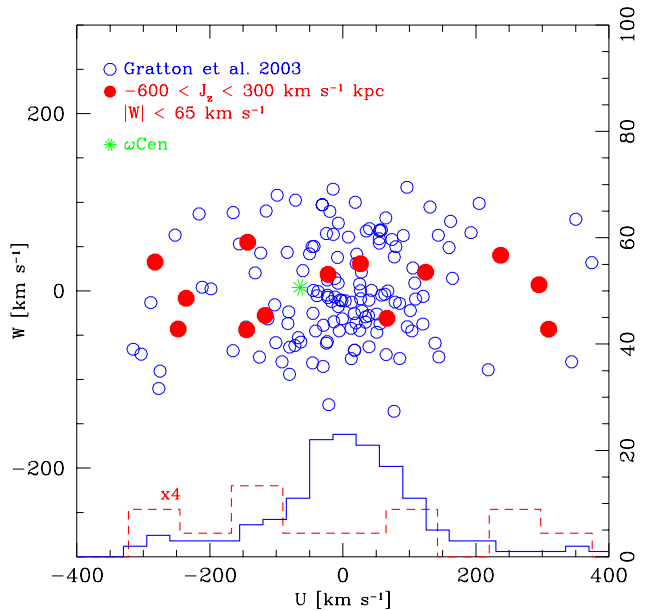}
\end{center}
\caption{
Vertical ($W$) versus radial ($U$) velocities of metal-poor stars in the Gratton
et al. (2003) compilation (open circles). Filled circles are used to highlight
$\omega$Cen group candidates. These are stars with angular momenta similar to
$\omega$Cen's (starred symbol), and with relatively low vertical velocity (see
conditions as labelled in the figure). The $U$ distribution of the $13$
candidate stars (dashed histogram, multiplied by 4 for clarity, scale on right
axis) differs markedly from that of the rest of the stars in compilation (solid
histogram). Only 3 candidate stars have $|U|< 100$ km s$^{-1}$, whereas $9$
would be expected from the overall distribution.
\label{figs:uw}}
\end{figure} 

Gratton et al. (2003, hereafter GCCLB) have compiled element-to-element abundance
ratio measurements for about $150$ stars in the solar neighbourhood with
accurate kinematical data, bringing together their own measurements with
literature data from the work of Nissen \& Schuster (1997), Fulbright (2000),
and Prochaska et al. (2000). The angular momentum distribution of these stars is
presented in Fig. \ref{figs:jzgrat}. The ``excess'' of stars in the B00
catalog corresponding to the $\omega$Cen and Arcturus groups noted in
Fig. \ref{figs:vphihist} are also clearly seen in this compilation,
underscoring the robustness of these kinematic features to various selection
techniques.

Following the motivation outlined in Section \ref{sec:numexp}, we analyze the Galactocentric
radial ($U$) and vertical ($W$) velocity\footnote{We use $U$, $V$, and $W$ to denote radial,
tangential and vertical velocities in Galactic coordinates, measured relative to the LSR.
$U$ is positive outwards, $V$ in the direction of the Sun's rotation, and $W$ towards the
North Galactic Pole. We assume that the Sun's ($U$,$V$,$W$) velocity is ($-9$,$12$,$7$) km
s$^{-1}$.} distribution of $\omega$Cen group stars in Fig. \ref{figs:uw}. Open circles
show the vertical and radial velocities of all stars in the GCCLB compilation, whereas the
filled circles highlight those with angular momenta in the ``peak'' associated with
$\omega$Cen in Fig. \ref{figs:jzgrat}, i.e., $J_z$ between $-600$ and $300$ km s$^{-1}$ kpc.
We choose to focus on those stars which, like $\omega$Cen, have their orbits roughly
confined to the disk, and restrict the candidate list further to stars with $|W| < 65$ km
s$^{-1}$ in order to weed out unrelated halo stars.

Thirteen stars are selected by these two criteria as candidate members of the group, and
their $U$ distribution is shown by the dashed histogram in Fig. \ref{figs:uw} (for
clarity, the number of stars in each bin is multiplied by four relative to the scale on the
right axis). The radial velocity distribution of these stars is in obvious contrast with
those of all stars in the GCCLB compilation, which is well described by a Gaussian of zero
mean and dispersion $125$ km s$^{-1}$ (solid histogram in Fig. \ref{figs:uw}), as expected
for stars selected preferentially near the apocenter of their orbits. The $\omega$Cen group
candidates, on the other hand, span a large symmetric range in $U$, from $-300$ to $300$ km
s$^{-1}$, with no obvious central tendency (i.e., few stars near apocenter) and a velocity
dispersion exceeding $200$ km s$^{-1}$. Indeed, only three out of thirteen candidates have
$|U|< 100$ km s$^{-1}$, whereas one would have expected three times as many had their $U$
distribution been similar to that of the whole ensemble. 

With the obvious caveat that the number of candidates is rather small, one is
tempted to interpret the broad, roughly symmetric $U$ distribution as evidence
that some of the candidate stars are part of a dynamically coherent group shed
by $\omega$Cen's dwarf prior to disruption. The apocenter of this group is
outside the solar circle, and therefore group stars are only seen moving towards
apocenter with high positive $U$ velocity or coming from the apocenter with high
negative $U$.

\begin{figure}
\begin{center}
\includegraphics[width=\linewidth,clip]{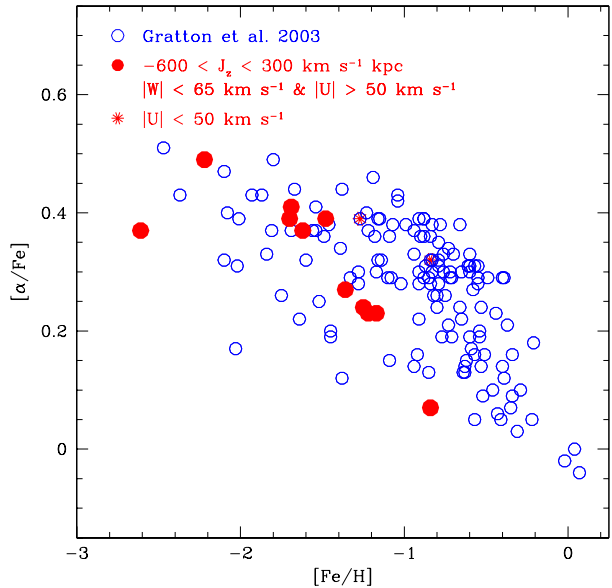}
\end{center}
\caption{Abundance ratios for stars in the Gratton et al. (2003) sample. Open circles
correspond to all stars in the sample, whereas filled circles indicate candidate
stars of the $\omega$Cen group. Note that these stars delineate a well defined
sequence in the abundance ratios that may be interpreted as resulting from a
protracted episode of star formation where a system self-enriched to a
metallicity of order one-tenth of solar. Two candidate stars deviate from the
main trend, and are highlighted with starred symbols. These two stars, however,
are kinematically distinct from the rest. Although they share the angular
momentum of $\omega$Cen, they are at the apocenter of their orbits ($|U|<50$ km
s$^{-1}$) in sharp contrast with the rest of $\omega$Cen group candidates, which
are moving through the solar neighbourhood to or from their apocenter at an
average radial velocity of $\sim 200$ km s$^{-1}$. These stars are unlikely to
be related to $\omega$Cen's progenitor.
\label{figs:fehafe}}
\end{figure}

Support for the common origin of this group comes from analyzing the abundances
of candidate stars. Fig. \ref{figs:fehafe} shows the abundance of $\alpha$
elements and iron for stars in the GCCLB sample. As is customary, the
$\alpha$-element abundance is expressed as the logarithm of its abundance
relative to Fe in solar units, [$\alpha$/Fe], and the iron abundance is given as
a logarithmic measure of its ratio to H also in solar units, [Fe/H].  Open
circles correspond to all stars in the GCCLB compilation, whereas the filled
circles indicate the $\omega$Cen group candidates, as in
Fig. \ref{figs:uw}. Remarkably, the $\omega$Cen group candidates are not
distributed at random amongst the sample, but rather trace a well-defined narrow
track in this plane, spanning a wide range in iron abundance
($-2.6<$ [Fe/H] $<-0.9$) which is consistent with the spread in metallicity
measured for individual stars in $\omega$Cen (Suntzeff \& Kraft 1996).

This well defined track in the [$\alpha$/Fe] versus [Fe/H] plane is
roughly consistent with simple one-zone self-enrichment models (see,
e.g., Matteucci \& Francois 1989), but where pollution by the
iron-rich ejecta of type Ia supernovae becomes important at lower
metallicity than for ``typical'' stars in the halo, where
[$\alpha$/Fe] drops precipitously for [Fe/H] $\gsim
-1.5$. Intriguingly, two candidate stars appear to deviate from this
trend, and are highlighted by starred symbols in
Fig. \ref{figs:fehafe}. This is consistent with the above
interpretation, as these two stars are dynamically quite distinct from
the rest; although they share the angular momentum of $\omega$Cen,
their low $U$ velocity ($|U|<50$ km s$^{-1}$) sets them apart from the
rest as unrelated stars on much more tightly bound orbits. 

The chemical coherence highlighted by the filled symbols in
Fig. \ref{figs:fehafe} is thus manifest only when stars of coherent
kinematics (i.e. similar $E$ and $J_z$) are considered part of the
group.  Overall, this pattern is consistent with that expected for a
satellite system that self-enriched to a metallicity of order
one-tenth of the Sun on a time scale somewhat longer than envisioned
for the bulk of metal-poor stars in the vicinity of the Sun.

We conclude that the chemical and dynamical coherence of stars in the
$\omega$Cen group is highly suggestive of a common origin, most likely
a single dwarf system that was dragged and disrupted into the disk of
the Galaxy early during its assembly history.

\section{Summary and Discussion}
\label{sec:conc}

We use numerical simulations to analyze the dynamical properties of tidal debris
stripped from a satellite on a highly eccentric orbit roughly coplanar with the
disk of the main galaxy upon disruption. Because of the small pericentric radius
of the orbit, the stellar component of the satellite is disrupted by the tidal
shocks that accompany the last few pericentric passages, splitting the bulk of
stars into coherent groups of common energy and angular momentum on eccentric
orbits roughly confined to the disk.

Observers located between the pericenter and apocenter of such groups would see
the debris as stars of common angular momentum but with a wide, symmetric (and
at times ``double-peaked'') distribution of Galactocentric radial velocities. A
group of stars with these characteristics stands out in catalogs of nearby
metal-poor stars and share the peculiar kinematics of the unusual globular
cluster $\omega$Cen. These stars orbit the Galaxy on retrograde orbits but
confined to relatively small vertical excursions from the plane of the disk, and
show extremely well-correlated element-to-element abundance ratios. They span a
range in [Fe/H] comparable to that of individual stars in $\omega$Cen, and
follow closely the [$\alpha$/Fe] versus [Fe/H] correlation expected for a simple
closed-box self-enrichment model.  The dynamical and chemical coherence of stars
in this group are highly suggestive of a common origin, which we ascribe to the
dwarf galaxy that brought $\omega$Cen into the Galactic disk.

The $\omega$Cen group thus shares some of the properties of the ``Arcturus
group''; a group of stars of moderate velocity dispersion and singular metal
abundance (Eggen 1971), that Navarro et al. (2004) have recently claimed to be
part of the debris from another past accretion event. However, the $\omega$Cen
and Arcturus groups also differ in a number of properties.  For example,
although Arcturus group members have negligible net radial motions and are thus
near the apocenter of their orbits, $\omega$Cen group stars are passing by the
solar neighbourhood at fairly high speeds to or from apocentric radii far
outside the solar circle.

In addition, most Arcturus group members have highly enhanced [$\alpha$/Fe]
ratios relative to solar, and are in this regard similar to the majority of
metal-poor stars in the solar neighbourhood (see Figure 3 of Navarro et
al. 2004).  On the other hand, the [$\alpha$/Fe] ratio of $\omega$Cen group
members approaches the solar value at relatively low metallicity ([$\alpha$/Fe]
$\sim 0.05$ at [Fe/H] $\sim -0.9$; see Fig. \ref{figs:fehafe}). Thus, several
stars in this group have unusually low [$\alpha$/Fe] enhancement relative to
typical halo stars of comparable metallicity. This is important, as it suggests
an ``extra-Galactic'' resolution for the puzzle surrounding stars with
anomalously low [$\alpha$/Fe] in the solar neighbourhood: namely, all such stars
may come from disrupted dwarfs (Carney et al. 1997; King 1997; Hanson et
al. 1998; Fulbright 2002). Indeed, low [$\alpha$/Fe] ratios are consistent with
those of stars in the dwarf satellites of the Milky Way (Draco; Sculptor; and
Sagittarius; see, e.g., Shetrone et al. 2001, 2003; Bonifacio et al.  2004; Venn
et al. 2004).

The prominence of the Arcturus and $\omega$Cen groups in samples of
metal-deficient stars (the two combined make up to $\sim 16\%$ of the
stars in the GCCLB sample) suggests that accretion events have
probably contributed a significant number of the metal-deficient stars
in the solar neighborhood. These two examples demonstrate that at
least some, and perhaps many, old stars on orbits confined to the disk
did not form {\it in situ}, but were brought into the Galaxy by
accretion events.

It is clearly important to search exhaustively for signatures of past
accretion events in the solar neighbourhood in order to quantify
properly the contribution of tidal debris to the inventory of
metal-poor stars in the Galaxy, as this may seriously affect the
validity and applicability of Galactic formation and evolution
models. For example, ELS-inspired interpretations of the correlation
between metallicity and rotational support for nearby metal-poor stars
in terms of a global collapse rest on the assumption that their orbits
have not evolved drastically over time{\footnote{We note that Chiba \&
Beers (2000) argue against the presence of such correlation in
reference to ELS' work. However, this applies only to the most metal
poor tail of the halo distribution, i.e., to stars with [Fe/H] $\lsim
-2.2$ (see also Bekki \& Chiba 2000 for further discussion). Our
comment, on the other hand, refers to all stars in the Beers et al. (2000)
sample, i.e., those with [Fe/H] $\lsim -0.6$.}}.

On the other hand, if most metal-poor stars in the solar neighbourhood
have been brought into the Galaxy by accretion events, a drastically
different scenario may be required to accommodate this
correlation. One possibility is that this may just reflect the
increasing efficiency of dynamical friction in circularizing the
orbits of more massive satellites---the most effective contributors of
stars relatively rich in metals.

Is it possible that the whole trend seen in Fig. \ref{figs:fehafe} is
caused by the overlap of stars stripped from numerous satellites, each
of which may have self-enriched roughly as a closed box prior to
accretion?  It would be premature to argue that the data is fully
consistent with such scenario, but nor is it obviously ruled out by
the data.

For example, the scarcity of metal-poor stars with low [$\alpha$/Fe] may be
explained by arguing that they originate in the smallest and most metal-poor of
all accreted satellites (i.e., the likes of Sculptor, Draco, or the progenitor
of $\omega$Cen). Overall, few stars are locked up in such small satellites, and
therefore it is not surprising that they contribute a relatively small fraction
of all the metal-deficient stars in the solar neighbourhood. In this
interpretation, the main trend in the [$\alpha$/Fe] versus [Fe/H] plane
(Fig. \ref{figs:fehafe}) is dominated by the enrichment process of the few
massive dwarfs whose debris dominates the metal-poor star counts in the solar
neighbourhood. One example would be the progenitor of the Arcturus group, which
probably self-enriched rapidly to [Fe/H]$\sim -0.5$ before disruption, and
therefore contributed mostly stars highly enhanced in [$\alpha$/Fe].

A number of critical enquiries should be pursued before one may elevate this
speculation into a working hypothesis to guide the interpretation of future
datasets. These enquiries will soon become possible, as the completion of
large-scale surveys such as those planned by the RAVE collaboration (Steinmetz
et al. 2003) or by GAIA (Perryman et al. 2001) promise to provide a
comprehensive appraisal of the importance of kinematical substructure in the
dynamics of the Galaxy.

Detailed abundance studies of stars selected from such surveys to
belong dynamically to either the Arcturus or $\omega$Cen groups may
corroborate the ``closed-box'' evolutionary sequence outlined for the
progenitors of these stellar groups. (We note, however, that this may
only apply to the low-mass end of accreted dwarfs, as large satellites
like, say, the LMC, may themselves have been built through the
assembly of smaller subunits.) The discovery of further dynamically
and chemically coherent groups associated with stars of anomalous
chemistry in the solar neighbourhood would be especially helpful to
confirm the extra-Galactic origin of such stars. 

Finally, ongoing studies of the kinematics of thick disks in external
galaxies (see, e.g., Dalcanton, Yoachim \& Bernstein 2004) hold the
promise of extending these studies beyond the unique environment of
the Milky Way and to unravel the cosmological role of accretion events
in the formation of the various galactic components.

\section*{Acknowledgments}

AM acknowledges support from the Comit\'e Mixto ESO-Chile and from the
Facultad de Ciencias F\'{\i}sicas y Matem\'aticas of the Universidad de Chile.
JFN is supported by Canada's NSERC, the Canadian Foundation for Innovation,
and the Alexander von Humboldt Foundation.

\end{document}